# A detailed review of blockchain and cryptocurrency

Nayak Bhatia
MBA Tech IT
MPSTME, NMIMS
Mumbai, India
nayakbhatia1969@gmail.com

Sanchit Bansal
MBA Tech IT
MPSTME, NMIMS
Mumbai, India
sanchitbansal225@gmail.com

Smit Desai
MBA Tech IT
MPSTME, NMIMS
Mumbai, India
smitdesai1001@gmail.com

*Abstract* – **Cryptocurrency is something that we have all heard about recently, most likely preceded by bitcoin, and how much its prices have boomed over the decade. These cryptocurrencies are actually based on blockchain, a secure datatype, and recently popular form of technology. This paper gives a detailed review about the concept of blockchain and its potential applications, especially elaborating on cryptocurrency, and it also contains a detailed case study of "blockchain Dubai".**

## I. INTRODUCTION

Blockchain Technology is a real-time ledger for recording financial transactions including but not limited to physical assets, contracts, and much more. It is a purely decentralized system or network, meaning that there is no single person or organization in charge of the entire network or chain. One more important feature is that it is open for all, which means everyone within the network can see the details of the transactions [1]. Every single block in the network is secured using the method of cryptography. This method involves a transaction signature by a private key and then further verified by a public key. If any data changes occur, the key or that unique signature becomes invalid, resulting in that particular block's exclusion from the network or chain.

Blockchain is also referred to as Distributed Ledger Technology (DLT), where the history of any transaction is stored in the form of decentralized and immutable blocks [2safe]. It is a type of database where the data can be recorded and distributed but not edited, making it unalterable.

Stuart Haber and W Scott Stornetta first described blockchain technology in the year 1991, but its first real-world application came into existence in 2009 with the launch of the first cryptocurrency 'Bitcoin.'

**Why is this technology described as "Blockchain"?**

A blockchain is essentially a type of chain that accumulates data in the form of groups, commonly referred to as blocks. Each block has a finite amount of storage and capacity, and due to that, when it is filled, it is linked to the previous filled block, forming a chain. Hence, the name "Blockchain."

Each block of the chain consists of three elements:
• Relevant information
• Hash
• Hash of a preceding block

Every time there is a transaction, a hash is generated, and the hash is in the form of a string made up of numbers and letters (e.g., 6U9P2).

**How hashing works?**

Hash is a type of connection where its input (numbers and letters) is converted to the fixed length of the encrypted output.

These hashes are dependent not only on the transaction but also on previous transaction hashes, so if we try to change a little bit in any of the transactions, a new hash will be generated. If someone tried to change any data in a blockchain, they would get caught quickly because of a massive change in the data. As a result, it is regarded as a safe technology [3].

This blockchain is distributed across multiple computers, with each computer having a copy of the blockchain. These machines are referred to as nodes. These nodes are in charge of checking the hashes to ensure that the transaction has not changed, and if the majority of nodes approve the transaction, it is written in the block.

These nodes form the blockchain's infrastructure, storing, spreading, and preserving its data. A complete node is a sort of computer equipment that copies the blockchain's transaction history. The most delicate part of the blockchain is that it updates itself every 10 minutes, which helps maintain its security.

## II. ADVANTAGES OF BLOCKCHAIN

The term "blockchain" has become increasingly popular among the general public. The rise in bitcoin prices sparked its fame. However, according to Google, it is currently leveling off. Blockchain has become a vital aspect for various fields as it provides with:

1. **Immutability:**
    The blockchain is unchangeable, allowing platforms requiring immutable features to make their

systems highly functional in a crowded market. Take, for example, the supply chain. Immutability allows businesses to assure that the packages are not harmed in transit. It is impossible to change the package information in any way without alerting the system and most likely rendering the data void.

2. **Transparency:**

Transparency is another critical factor that makes blockchain vital. Of the many different forms of blockchain to exist, public blockchain gives transparency. It is pretty beneficial for a variety of purposes in our current society, e.g., conducting elections. Businesses can also utilize it to ensure that end-users interact with processes in full or partial transparency.

3. **Efficiency:**

Increased efficiency is also one of the reasons for the significance of the blockchain. Society nowadays relies more on efficiency than on cost-effectiveness, and the enhanced security, removal of intermediaries, & better processes of blockchain make it highly efficient. Transactions, especially international, take seconds rather than weeks to complete.

4. **Security:**

Blockchain adds an extra layer of protection to the data stored on the network by using cryptography, providing us with superior security compared to the previous systems. Cryptography is a method of securing data and systems on the blockchain network by employing complicated mathematical functions and formulae.

Furthermore, each block on the network has its own hash, ensuring that no data can be falsified or altered by hostile forces or attackers.

Most importantly, blockchain is a decentralized form of database, meaning that no person or organization has control over it. The ledgers are open for all to view and add on to (considering the access and permissions).

## III. APPLICATIONS AND POTENTIAL USES

We should know that the blockchain has grown in importance in recent years, not only in the crypto market but also in various other fields like finance, banking & healthcare. Its most popular applications and potential uses are:

- **Cryptocurrency**

Cryptocurrency, commonly abbreviated as crypto, is probably one of the most popular uses of the blockchain. A decentralized form of currency with no governing body, cryptocurrency has rapidly gained popularity. The transactions are generally stored in a blockchain and translated by cryptography.

- **NFTs**

NFTs are Non-Fungible Tokens, i.e., they are unique and cannot be interchanged. They can be anything digital, ranging from music, art, or even drawings, to an AI. For example, a dollar note is more or less the same; trade one for another, and one still has the same thing, both physically and monetarily. Whereas an NFT is unique, trade one for another, and you get something entirely else. Most NFTs work on the Ethereum blockchain, which is functional for a cryptocurrency (ETH), but it also serves NFTs.

Notably, some of the other blockchains have also implemented the use of NFTs. A few famous examples are "The Sandbox," priced at $595 million, "DigiByte," priced at $927 million, and "Flow," priced at a whopping $1,212 million!

- **Banking and Finance**

Blockchain has the potential to revolutionize the banking and finance industries. When it comes to smart contracts, digital financial institutions stand to benefit the most. Digital assets, programmable money, and smart contracts provide the benefits. There are numerous applications in the banking and financial industry, and in recent years, trade finance blockchain has received much interest.

- **Healthcare**

Healthcare is another sector where blockchain will play an important role. Currently, healthcare is suffering from a lack of unity. Patients need to carry their documents, and hospitals put records in separate silos, which takes time to be retrieved when needed [4].

- **COVID Vaccine model**

One in four of the COVID-19 vaccines were initially being wasted because of poor distribution and proper packaging. The IBM blockchain developed a COVID vaccine distribution model, which would maintain and ensure a proper record of all the successful vaccinations, which helped in proper distribution and real-time visibility & enhanced ability to respond to problems.

## IV. CRYPTOCURRENCY

Cryptocurrency is a virtual or digital currency. The word "Cryptocurrency" came into existence or was invented in 2008 basically by an unknown group of people using the pseudonym Satoshi Nakamoto.

Blockchain Technology and Cryptocurrency are interconnected as the method of Cryptography is used to secure cryptocurrency, and the records of all transactions are stored on blockchains.

Cryptocurrencies work as a medium of exchange wherein individual coin ownership records are stored in a ledger in the form of a database. By distributing its functions across a network, blockchain technology enables cryptocurrency to work without a central host or power. One crucial thing in cryptocurrency that happens with the help of blockchain is that the currency is decentralized, and it also gives real-time access to people, which makes it more reliable to the younger generations.

Cryptocurrency and blockchain technology together form an innovative and promising technology as it reduces risk, eliminates financial fraud, thereby increasing the transparency of a system as a whole. Bitcoin, Ethereum, Dogecoin, etc., are famous examples from over 5000 different cryptocurrencies. Currently, China holds a monopoly over cryptocurrency mining.

## V. POPULARITY GAIN

Cryptocurrency has always been very volatile in terms of money. The total cryptocurrency market capitalization peaked in 2021 at about $2.4 trillion, up from around $200 billion in previous years. Due to cryptocurrency being backed by many influential people like Elon Musk, Snoop Dogg, Bill Gates, and Indian rapper, Raftaar.

Cryptocurrency is becoming a much more reliable option as compared to centralized currencies in the long run. To understand this, we need to study the history of money.

Before rupees and dollars, gold used to be the currency. Since it was heavy to carry around, and a huge safety risk, one man started a safe depositing place for the gold and gave out receipts. Now, instead of going to the safe place before any transaction, the people realized it was easier to trade the receipts, and that is how the currencies were born. These safe places (banks) now started lending the receipts and soon lent substantially more than the gold they actually had with them. Once the population caught on, the banks ran away and were then bailed out by the government. Since then, this has been the simulation; every time the bank is in trouble, the government (central controller of the currency) comes to its rescue.

Cryptocurrencies do not have such shortcomings, as they are decentralized, and no one person has control over it.

## VI. POPULAR CRYPTOCURRENCIES

**Bitcoin**

Bitcoin is a decentralized digital currency and is most known in the form of cryptocurrency. Bitcoin is also the first cryptocurrency invented in 2009 by a group of unknown people. It is a network that runs using blockchain technology. Bitcoins are created as a reward for a process known as mining. Bitcoin Miners run or solve complex puzzles to secure a group of transactions, also known as blocks. On getting successful, these secured blocks are added to the blockchain, and the miners are rewarded with a small number of bitcoins. Miners have become more advanced and sophisticated, using advanced machinery to increase the speed of mining. Cryptocurrencies, including Bitcoin, have a negative impact on the environment. More and more mining activities lead to additional energy consumed by computers as they produce heat and need to be kept cool.

Bitcoin has the highest market share among all the cryptocurrencies, to be precise, 44%. It is also first based on market cap among all the cryptocurrencies.

1 BTC = 40,46,407.74 INR
Total units of BTC = 21 billion
Current market capitalization = $632.7 billion

**Dogecoin**

Dogecoin is a cryptocurrency invented or created by engineers Billy Markus and Jackson Palmer initially created as a payment system, now a legitimate investment prospect. Dogecoin works on the principle of blockchain technology similarly to bitcoin and Ethereum. Dogecoin's blockchain network uses cryptography to keep transactions safe and secure. Dogecoin, when compared to bitcoin, is easier to get for miners. Currently, the market capitalization of Dogecoin represents about 1.4% of total crypto market capitalization.

1 DOGE = 16.87 INR
Total units of DOGE = 131.4 billion
Current market capitalization = $30.58 billion

**Ethereum**

Ethereum is a decentralized, open-source blockchain with smart functionality [5]. It comes second to bitcoin in terms of market capitalization. Ethereum has its own associated cryptocurrency, also known as ETH. The users of this network can create, publish, and monetize the use of applications. Ethereum has established itself as a programmable blockchain wherein people can go through financial services, play games, and use paid apps using Ethereum cryptocurrency. This is also beneficial as it prevents financial fraud and thefts. Ethereum has the second-highest market share of 14% after bitcoin.

1 ETH = 2,69,891.99 INR
Total units of ETH = 96 million
Current market capitalization = $500 billion

## Litecoin

Litecoin is an open-source, peer-to-peer cryptocurrency invented in 2011 by Litecoin Core Development Team. Litecoin is very similar to bitcoin when it comes to the technical part of the network. Charlie Lee, the core member of the development team, also stated that Litecoin is the "lite version of bitcoin." Unlike other cryptocurrencies, the total supply of Litecoin is definite. There will never be more than 84 million Litecoin in circulation [6]. There are rewards for mining Litecoin. Each time a miner successfully gets a block or verifies it, they are rewarded with 12.5 Litecoin, and the Litecoin awarded for this task will reduce with time. Litecoin stands at 15$^{th}$ rank based on market cap

  1 LTC = 13,442.00 INR
  Total units of LTC = 2.3 million
  Current market capitalization = $8.77 billion

## Binance

Binance is the largest cryptocurrency exchange in terms of the daily trading volume of cryptos [7]. It was founded in 2017 in Hong Kong. The name Binance is basically a combination of "Bitcoin" and "finance," which assures the exchange of cryptocurrencies providing a safe and secure platform. Binance provides a platform for users to store their electronic funds. Binance has its own blockchain-based coin, also called a Binance coin.

Binance also offers many other services like Binance LaunchPad, Binance smart pool, Binance Visa Card, etc.

  1 BNB = 32,075.86 Indian Rupees
  Total units of BNB = 200 million
  Current market capitalization = $72.2 billion

## Ripple (XRP)

Ripple is a cryptocurrency and also a digital payment system for financial transactions, released in 2012. Ripple is the name of the blockchain network and company, whereas XRP is a cryptocurrency token [8]. Ripple uses less energy compared to bitcoin, and the transactions are also less compared to bitcoin. Ripple (XRP) ranks sixth based on market capitalization. It is open-source software and allows easy transfer and exchange of currencies, including Euros, Dollars, Yen, and cryptocurrencies like bitcoin. The current market share of ripple is at 2.1%.

  1 XRP = 40,46,407.74 INR
  Total units of XRP = 45,404 billion
  Current market capitalization = $10 billion

## Chainlink (LINK)

It is a decentralized Oracle network, first announced in 2017 and implemented on the Ethereum blockchain in 2019. Sergey Nazarov and Steve Ellis were the first to create or invent it.

Chainlink is a blockchain abstraction that allows connected smart contracts to be operated globally. While blockchain cannot connect to external applications, chainlink's decentralized Oracle can.

  1 LINK = 2,265.21 INR
  Total units of LINK = 1 billion
  Current market capitalization = $13.5 billion

## TETHER (USDT)

Tether was the first cryptocurrency to create a stable coin group. Coins were created with the goal of reducing market volatility, and we all know how volatile the cryptocurrency market is; even the major players like bitcoin have an extremely volatile and unpredictable market. Tether and other stable currencies were created to help level out the market's excessive price volatility.

John Betts was the creator of this cryptocurrency, which was introduced in 2014. Tether is now the 5th biggest cryptocurrency by market capitalization as of September 2021. The Tether is directly tied with the US dollar

  1 USDT = 75.15 INR
  Total units of USDT = 69.3 billion
  Current market capitalization = $68.3 billion

## Cardona (ADA)

Cardona is another decentralized cryptocurrency that allows for peer-to-peer transactions. Cardona was established in 2015 by Charles Hoskinson, who was also a founding member of the Ethereum project. However, he quit that team owing to a disagreement on Ethereum's orientation and eventually formed the Cardona.

After researching and testing with numerous research papers, the Cardona team was able to build the blockchain. Cardona was made feasible by a team of scholars that combed through more than 90 publications on the blockchain issue and its various subtopics.

  1 ADA = Rs 187.83
  Total units of ADA = 45 billion
  Current market capitalization = $71 billion

## Monero (XMR)

Monero is an open-source cryptocurrency founded back in April 2014, and the person who founded this currency was Nicolas van Saberhagen. After the launch of the currency, it soon became popular among crypto aficionados and its community member. The main reason for its popularity was its anonymity as it provides complete privacy, which makes it untraceable, which means a transaction cannot be traced because of its exceptional security mechanisms [9].

As a result, it has a negative image since it has been tied to various illegal acts worldwide. The technique used to provide privacy to the coin is " Ring signatures."

  1 XMR = 21,134.27 INR
  Total units of XMR = 18.4 million
  Current market capitalization = $245 million

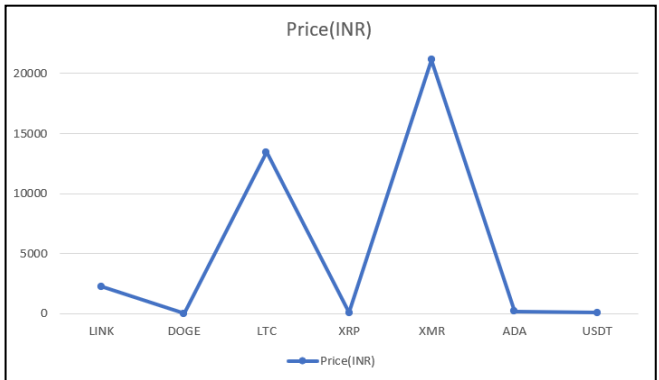

*The prices of BTC and ETH have been purposefully omitted so as not to deviate the graph drastically.

## VII. CASE STUDY – BLOCKCHAIN DUBAI

Dubai is becoming the Blockchain capital of the world [10]. The UAE government has implemented this innovative and efficient flow of network in conducting their transactions. Launching Dubai Blockchain Strategy aims to make Dubai powered with blockchain technology, thereby making it the happiest country in the world. If you want to buy a house someplace, you need to go around to different agencies and entities with lots of documents which becomes chaotic at times and is very energy and time-consuming. In Dubai, with the help of blockchain technology, one can access information directly from a secure safe digital network. Each piece of documentation, including the sales and purchase agreement, possession letter, allotment letter, etc. It is stored on a digital network in the form of blocks. This will eventually save time, energy and will be more efficient in this modern world.

Moreover, this block of information cannot be altered or changed. This block can contain every single piece of information, including passport details, bank details, etc. This seamless or smooth way of transactions and storing information offers trust and transparency. Dubai will be one of the first countries to do all of its financial transactions using Blockchain Technology.

**Vehicle History Blockchain Project**

The government of UAE, along with the transport authority, aims to create a vehicle lifecycle management project with the help of Blockchain Technology. The automobile industry is showing considerable growth around the world. This urges people to have a transparent and trusted process of the lifecycle of a vehicle. Using Blockchain Technology, the car dealers, manufacturers, along with buyers and sellers, will have a clear picture of vehicles' lifecycle right from the manufacturing plant to the scrap yard. This technology will help to provide transparency in matters involving vehicle transactions. This will also include ownership of the vehicles, their sale, and also its accident history. This provides a good ownership experience which leads to customer satisfaction.

## VIII. CONCLUSION

Blockchain is a relatively new and growing technology, with many possible applications, ranging from cybersecurity & banking, to medical tracking. The most popular application, cryptocurrency, the decentralized money is the path to our future, if utilized properly.